\documentclass[%
 reprint,
 amsmath,amssymb,
 aps,
 pra,
 longbibliography,
]{revtex4-1}

\usepackage{graphicx}
\usepackage{dcolumn}
\usepackage{bm}
\usepackage{natbib}
\usepackage{hyperref}
\usepackage{siunitx}
\usepackage[caption=false]{subfig}
\usepackage{booktabs}
\usepackage[export]{adjustbox}
\usepackage{multirow}
\usepackage{physics}
\usepackage{accents}

\begin{document}


\title{Fast entangling gates in long ion chains}

\author{Zain Mehdi}
 \email{zain.mehdi@anu.edu.au}%
 \affiliation{Department of Quantum Science, Research School of Physics, Australian National University}%
\author{Alexander K. Ratcliffe}
 \affiliation{Department of Quantum Science, Research School of Physics, Australian National University}%
\author{Joseph J. Hope}
 \affiliation{Department of Quantum Science, Research School of Physics, Australian National University}%

\date{\today}

\begin{abstract}
We present a model for implementing fast entangling gates (${\sim}1~\mu$s) with ultra-fast pulses in arbitrarily long ion chains, that requires low numbers of pulses and can be implemented with laser repetition rates well within experimental capability. We demonstrate that we are able to optimise pulse sequences that have theoretical fidelities above $99.99\%$ in arbitrarily long ion-chains, for laser repetition rates on the order of $100-300$~MHz. Notably, we find higher repetition rates are not required for gates in longer ion chains, which is in contrast to scaling analyses with other gate schemes. When pulse imperfections are considered in our calculations, we find that achievable gate fidelity is independent of the number of ions in the chain. We also show that pulse control requirements do not scale up with the number of ions.  We find that population transfer efficiencies of above $99.9\%$ from individual ultra-fast pulses is the threshold for realising high-fidelity gates, which may be achievable in near-future experiments.
\end{abstract}


\maketitle


\section{\label{sec:introduction}Introduction}
Trapped ion platforms are a promising platform for realising noisy intermediate-scale quantum (NISQ) computers \cite{Wineland1998,Haffner2008}. While there has been significant progress in demonstrating high-fidelity control for small numbers of qubits \cite{bermudez_PRX_Review}, scaling up trapped ion processors without slowing down gate speeds remains an open challenge. \par

High-fidelity entangling gates have been achieved using bichromatic light fields tuned near the motional sidebands to drive state-dependent trajectories through phase space \cite{Roos_2008,Molmer1999,Sorensen1999,Sorensen2000}. The M\o{}lmer S\o{}rensen (MS) mechanism requires a gate time significantly longer than the motional dynamics of the ions, in order to resolve the motional sideband transitions. 
Furthermore, as the number of ions $N$ in the trap is increased, the ion-mode coupling decreases and the axial trap frequency must be reduced to avoid buckling of the ion chain. If axial modes are used, both of these factors lead to a slower timescale for sideband resolution and thus longer gate times for gates in longer ion chains. This strongly limits the number of gates that can be performed before decoherence.\par
Outside of this `sideband-resolved' regime, multiple motional modes are excited by the ion-light interaction. Fast gate mechanisms use sequences of ultra-fast broadband laser pulses or amplitude-modulated continuous pulses to control the trajectories of each of these modes, and realise geometric phase gates \cite{Garcia-Ripoll2003,Garcia-Ripoll2005,Zhu2006,Steane2014}. Continuous-pulse fast gates have recently been demonstrated by Sch\"{a}fer \textit{et al}.~\cite{Schafer2018}, who demonstrated a high-fidelity ($99.8\%$) $1.6~\mu$s gate in a two-ion system, as well as a lower-fidelity (${\sim}60\%$)  $480~$ns gate. Fast gates with ultra-fast pulses have been experimentally demonstrated by Wong-Campos \textit{et al.}, who used the gate mechanism to prepare
an entangled Bell-state \cite{Wong-Campos2017}. In addition, there are several other groups developing experimental implementations of pulsed fast gates \cite{Hussain2016,Heinrich2019}. Ultra-fast pulses have also been used to demonstrate single-qubit control \cite{Campbell2010}. To date, fast gates have only been experimentally realised in two-ion systems. Developing fast gate schemes for many-ion systems is a key focus of this manuscript.  \par
The scaling behaviour of pulsed fast gates in large ion crystals has been theoretically studied in recent years \cite{Bentley2015,Taylor2017,Ratcliffe2018}. For chains of $10$ or more ions confined in a Paul trap, existing gate schemes require repetition rates of order $10-20$~GHz for implementing high-fidelity fast gates \cite{Bentley2015}. In microtrap arrays, fast gates scale more favourably, with pulse sequences optimised for gates in two-ion systems remaining robust when applied to arbitrarily large ion crystals \cite{Ratcliffe2018,Mehdi2020}. While this means that much lower repetition rates are required (${\sim}1~$GHz for gates on the same timescale as the motional frequencies), large number of pulses are typically required to entangle ions that are separated by ${\sim}100~\mu$m. We have previously identified that pulse errors are likely to be the dominant source of errors in fast-gates \cite{Gale2020}, and thus while fast gates in microtrap arrays are promising for their scalability, their experimental realisation requires improvement of pulse control from current experiments by several orders of magnitude. \par

In this manuscript we show that recently developed gate schemes can allow scalable processing in existing linear traps and with lasers repetition rates well within current experimental capabilities.  We find gate solutions that are faster than the trap period that can be implemented in long chains of ions with far fewer laser pulses than previous schemes. Notably, we find that pulse control requirements do not become restrictive as the number of ions increases. In fact we find that gates in arbitrarily long ion chains require approximately the same level of laser control as gates in two-ion systems. This result opens up a new pathway to scaling up trapped ion quantum computers without slowing down computation. 

\section{Fast gate formalism}

Pulsed fast gate schemes are composed of state-dependent momentum kicks (SDKs) from pairs of counter-propagating $\pi$-pulses, interspersed with periods of free evolution of the ions. These pulses are resonant with the electronic transition of the ions and can be described by the following interaction picture Hamiltonian:
\begin{equation}
    \label{eq:pulse_hamiltonian}
    H_I =  \sum_{j=1}^2\frac{\hbar\Omega(t)}{2}\left(\sigma^j_+ e^{i(kx^j+\phi)}+\sigma^j_-e^{-i(kx^j+\phi)}\right) \,,
\end{equation}
where $x^i$ is the deviation of the $i$-th ion from its equilibrium, $k$ and $\phi$ are the laser wavenumber and phase, and $\Omega(t)$ is the Rabi rate (satisfying $\int^t_0 \Omega(\tau)d\tau=\pi$). By expanding the ion's position into the mode basis, $k x^j=\sum_m b^j_m \eta_m (a_m+a^\dag_m)$, the resulting unitary can be written in terms of the mode displacement operator $D_m(\alpha)=e^{\alpha a_m^\dag-\alpha^* a_m}$:
\begin{equation}
    \label{eq:pulse_unitary}
    U_\pi=\prod_{j=1}^2\left(\sigma^j_+e^{-i\phi}\prod_m D_m(i b_m^j\eta_m) + \text{h.c.} \right) \,,
\end{equation}
where $b_m^j$ is the ion-mode coupling between the $j$-th ion and the $m$-th motional mode. The $m$-th motional mode has an associated frequency $\omega_m$ and Lamb-Dicke parameter  $\eta_m = k \sqrt{\hbar/2M\omega_m}$. A SDK can be built by applying a second $\pi$-pulse that is counter-propagating ($k{\rightarrow}-k$ or equivalently $b^j_m{\rightarrow}-b^j_m$). Assuming the two pulses are split from a single large pulse so the laser phase $\phi$ perfectly cancels, the SDK unitary can be expressed:
\begin{equation}
    \label{eq:SDK_unitary}
    U_\text{SDK} = \prod_{j=1}^2\prod_m D_m\left(-2i\eta_m b^j_m  \sigma_z^j\right) \,.
\end{equation}
The action of a single SDK on a coherent motional state can be understood as creating a cat state with $|\alpha| = 2\eta_mb^j_m$ in each of the $m$ motional modes. 
The unitary for the gate can be expressed as $N_p$ SDKs with free evolution between kicks:
\begin{equation}
    \label{eq:gate_unitary}
    U_\text{gate}=\prod_{k=1}^{N_p} U_\text{SDK}e^{-i\sum_m \omega_m \delta t_k a^\dag_m a_m} \,,
\end{equation}
where $\delta t_k$ is the time between the $(k{-}1)$-th and $k$-th SDK. In the ideal gate, a geometric phase gate is implemented on two ion-qubits indexed $\mu$ and $\nu$ with unitary $U_\text{id}=\exp(i\frac{\pi}{4}\sigma_z^\mu\sigma_z^\nu)$, with all motional modes decoupled from the ions electronic states by the end of the gate. 

\section{\label{sec:results}Model and results}

Previous analyses have identified that when the gate time $T_G$ is much faster than the motional dynamics ($T_G \ll 2\pi/\omega_t$), only the motions of local ions are affected by the gate, leaving the rest of the ion chain untouched \cite{Zhu2006,Bentley2015}. We exploit this in our model; as the axial frequency is reduced to accommodate longer ion chains, the motional dynamics slow down and thus distant ions are increasingly unlikely to be affected by a gate with a constant speed. This approach requires all gates to be between neighbouring ions; non-local gates can be implemented with the use of a series of local SWAP operations \cite{Taylor2017}. A consequence of this is that intermediate-scale quantum computations that require many non-local qubit operations may be inaccessible to this model. For such computations it may be preferable to utilise a multi-dimensional microtrap array, where fewer SWAP operations will be required to connect non-neighbouring ions \cite{Mehdi2020}. \par 

In our model we consider a chain of $^{40}\text{Ca}^{+}$ ions in a Paul trap with fixed radial frequency $\omega_r/2\pi=5~$MHz, and variable axial trap frequency $\omega_t = \omega_r/0.65N^{0.865}$ which is sufficient to prevent buckling of a chain of $N$ ions \cite{Schiffer93_buckling}. We consider ultra-fast laser pulses resonant on the \SI{393}{\nano\meter} $S_{1/2}\rightarrow P_{3/2}$ transition for the SDKs. We take the light field to be oriented down the length of the ion chain, such that only the axial modes are coupled in to the motion of the ions. We assume that all but the two target ions are shelved in states far from resonance with the laser. We will discuss the details and implications of such shelving later in this manuscript. \par
\begin{figure}[t!]
    \centering
    \includegraphics[width=1.0\columnwidth]{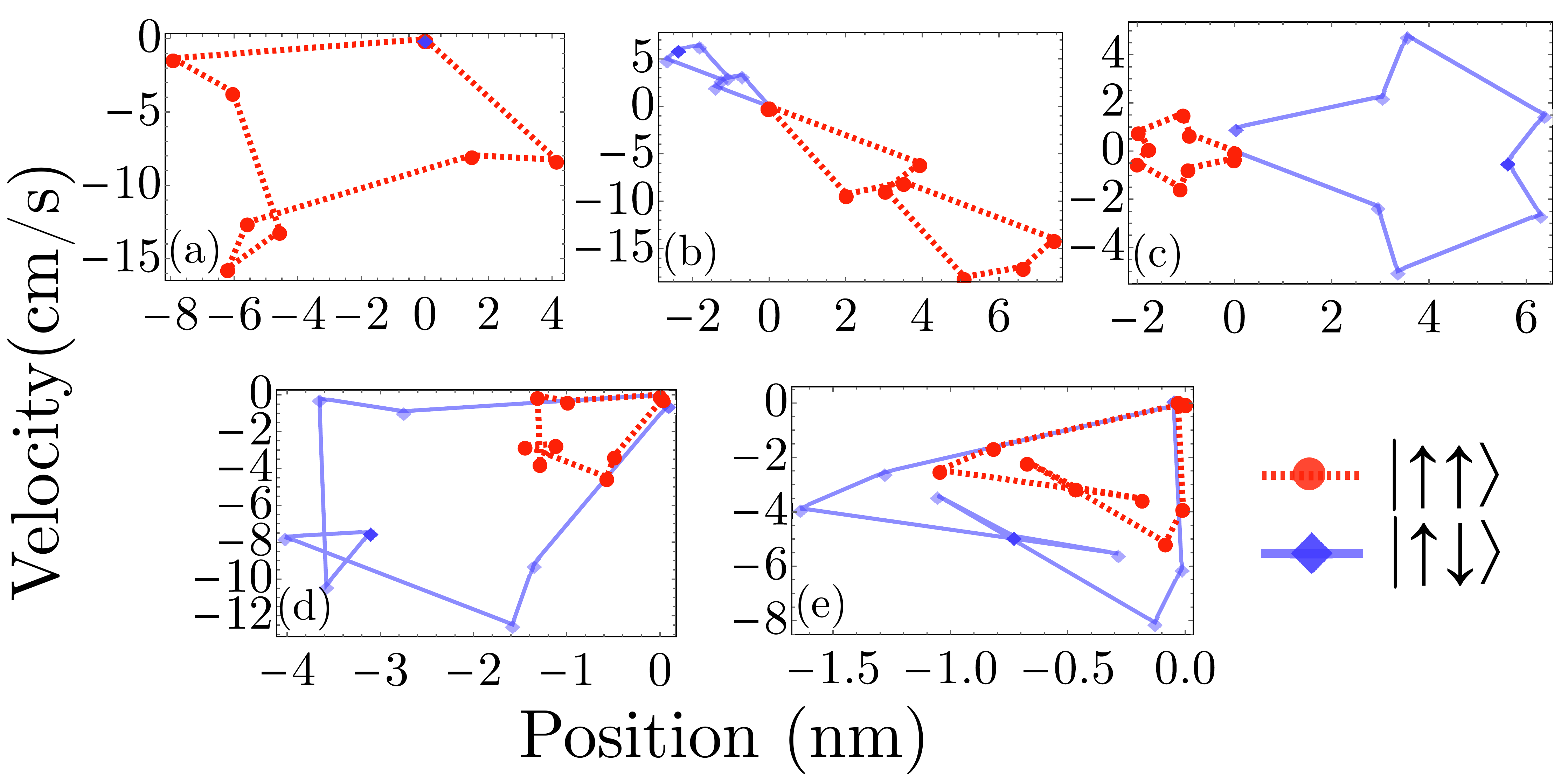}
    \caption{Simulated phase-space trajectories of the axial modes of a $N=5$ ion chain during gate with $16$ SDKs, for the two-qubit states $\ket{\uparrow\uparrow}$ and $\ket{\uparrow\downarrow}$. In the gate scheme, individual SDKs are grouped together so that some kicks can be made larger than others. The accumulated entangling phase is proportional to differences in areas enclosed by the $\ket{\uparrow\uparrow}$ and $\ket{\uparrow\downarrow}$ trajectories, summed for each motional mode. Here we represent each mode in position-velocity phase-space, rotating at its normal frequency: (a) $1.9$~MHz, (b) $3.3$~MHz, (c) $4.6$~MHz, (d) $5.6$~MHz, and (e) $7.0$~MHz. It is apparent that some of the trajectories do not close perfectly, which contributes to a non-zero infidelity of ${\sim}5\times10^{-5}$. }
    \label{fig:PStraj}
\end{figure}

We perform optimisation of the number of SDKs and their timings in order to minimise the state-averaged infidelity of the unitary \eqref{eq:gate_unitary} with respect to the ideal unitary $U_\text{id}$, following the two-step procedure outlined in Ref.~\cite{Gale2020}. In the first step, we consider a family of gate schemes where large state-dependent momentum kicks occur at regular time intervals. We optimise the magnitude of the kicks to minimise the gate infidelity, with the constraint that the kick sizes are integer multiples of $2\hbar k$. In the second step, we decompose each of these kicks into $2\hbar k$ SDKs from pulse-pairs separated by the laser repetition period. We then optimise the timings between SDKs on a discrete grid specified by the repetition rate. This second stage of optimisation is done using an ordinary differential equation (ODE) description of the gate dynamics to calculate the accumulated phase and residual coupling to the motional modes. For efficient integration of the ODEs in large ion chains, we truncate the Coulomb potential to second order and convert to a normal-mode basis where the ODEs decouple. Further details of our fidelity calculations, and further discussion of our optimisation approach can be found in Appendix \ref{Append:OptDetails}. The phase-space trajectories of an exemplary gate optimised for system of $N=5$ ions are shown in Fig.~\ref{fig:PStraj}. \par
To ensure these results are experimentally realistic, we include the effect of pulse errors when calculating fidelity after optimisation. 
We have previously shown that pulse errors reduce the fidelity as $F \simeq |1-N_p \epsilon|^2 F_0$ \cite{Gale2020}. Here $F_0$ is the ideal fidelity ($\epsilon=0$) and $\epsilon$ is a characteristic error in the population transfer of a single ultra-fast $\pi$ pulse. This expression is based on a worst-case analysis and is truncated to second order in $\epsilon$, which is accurate for $N_p\epsilon\ll1$. The pulse errors place strong constraints on the maximum number of SDKs in the gate, and thus we perform optimisations with tight bounds on the total number of SDKs allowed in the gates. Optimal pulse sequences are selected after pulse errors are included in the estimated gate fidelity.
\par 
Restricting the number of pulses allowed in any given gate has the consequence of limiting the achievable gate speed for reasonable experimental repetition rates; we find that gate times between $0.7-1.2~\mu$s are feasible with minimal numbers of pulses and repetition rates between $100$~MHz$-1$~GHz. In terms of trap units, these gate times are roughly $4\times(2\pi/\omega_t)$ for two-ion systems, and up to $0.2\times(2\pi/\omega_t)$ in $N=100$ ion chains in our model. While we have focused on gate times around $1~\mu$s in this manuscript, we note that our optimisation protocol allows us to find pulse sequences for much faster gates. However, faster gates require larger numbers of ultra-fast pulses, thus increasing the sensitivity of the gate fidelity to pulse imperfections. The pulse control requirements for high-fidelity gates faster than ${\sim}0.7~\mu$s will almost certainly be beyond reach of current or near-future experiments.
\par

\begin{figure*}[t!]
    \centering
    \includegraphics[width=0.95\textwidth]{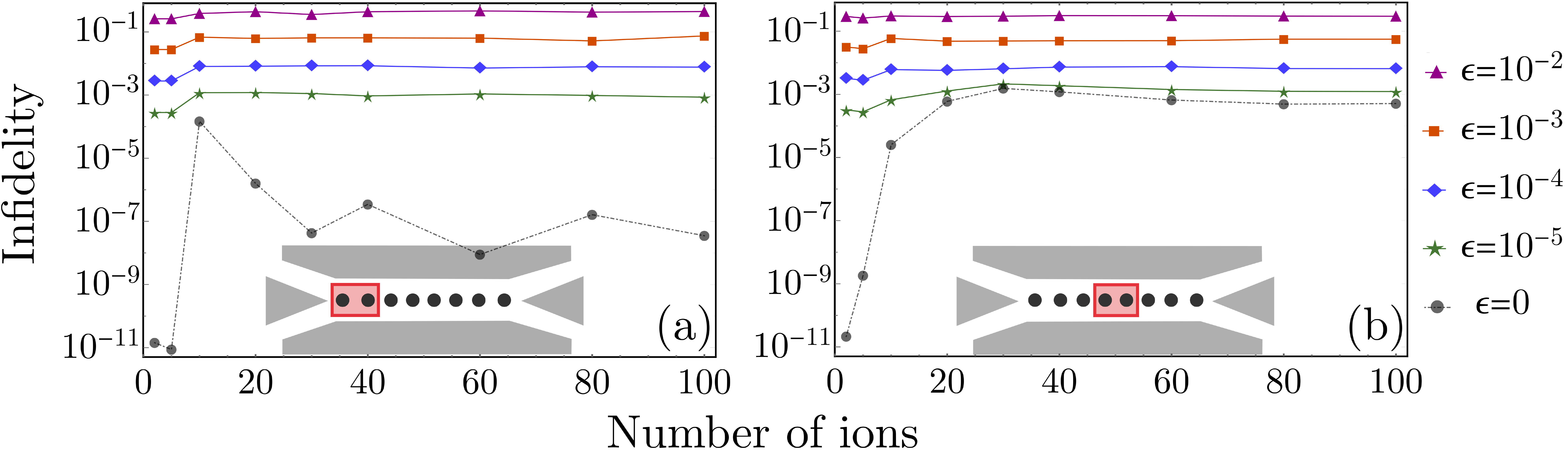}
    \caption{Infidelity of gates optimised for different numbers of ions in a trap, between two ions (a) on the edge of the ion chain and (b) in the middle of the chain. Each gate is optimised for a laser repetition rate of $300$MHz, and has a gate time between $0.75-1.2~\mu$s after optimisation. The effect of imperfect SDKs due pulse area errors is included in optimisation, with $\epsilon$ giving the characteristic population transfer error from a single $\pi$ pulse. For error rates of $\epsilon=10^{-5}$ or lower, fidelities above $99.99\%$ are achievable between neighbouring ions in arbitrarily long ion chains.}
    \label{fig:Scaling_300MHz}
\end{figure*}
\begin{figure}[t!]
    \centering
    \includegraphics[width=0.95\columnwidth]{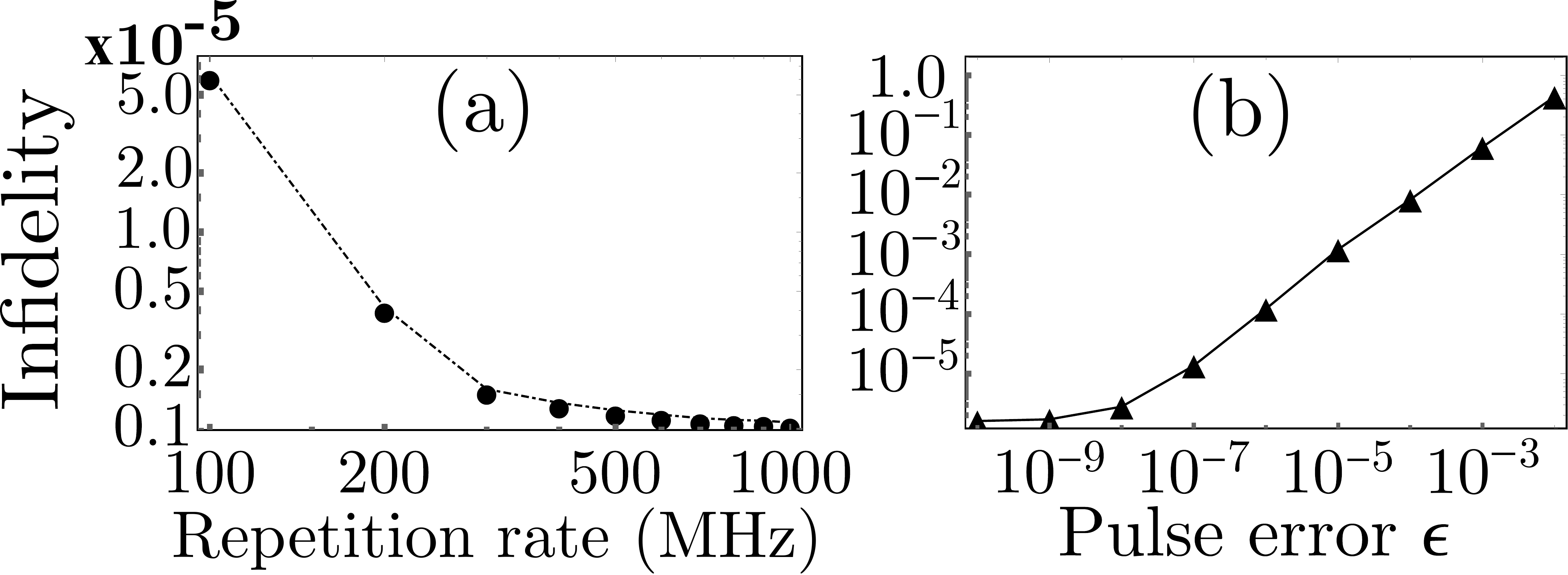}
    \caption{Infidelities of gates between two ions on the edge of a $N=20$ ion chain, as a function of (a) repetition rate, and (b) pulse error. (a) Infidelities are shown assuming perfect pulses ($\epsilon=0$) for gates optimised for different repetition rates. For repetition rates above about $300$~MHz, fidelity does not significantly improve. (b) Infidelities of gates optimised for a $300$~MHz repetition rate laser, with pulse imperfections (characterised by typical population transfer error $\epsilon$) included in the calculation. We find that achievable infidelity in long ion chains is approximately $1-F\sim 10^{2}\epsilon$. In the limit of perfect pulses ($\epsilon\rightarrow 0$), the infidelity approaches the ideal value $1-F_0 \sim 10^{-6}$.    }
    \label{fig:inf_vs_params}
\end{figure}
In Figure \ref{fig:Scaling_300MHz}, we show the results of gate optimisations in this model for ion chains of different lengths. We find that even in chains of up to $100$ ions the optimisation is able to find high-fidelity (above $99.9\%$) gate solutions with $T_G \simeq 1\mu$s. 
We observe a difference in ideal gate fidelities ($\epsilon=0$) for gates in different locations in the chain: optimised gates for ions at the edge of the chain are best placed, and ions in the middle of the chain are worst placed, with fidelities of about $1-10^{-7}$ and $1-10^{-3}$ in the large $N$ limit, respectively.
Fig.~\ref{fig:Scaling_300MHz} also shows that realistic gate fidelity is roughly independent of the number of ions in the chain. In Fig.~\ref{fig:inf_vs_params} we show that the achievable infidelity is roughly $1-F\sim10^2\epsilon$; thus to realise high-fidelities of above $99\%$ pulse errors of about $\epsilon=10^{-4}$ are required. 
We discuss the feasibility of achieving this level of pulse control in the following section. \par

Notably these results are all for gates optimised for a repetition rate of $300$~MHz, which is well within experimental feasibility \cite{Hussain2016,Heinrich2019}. In Figure \ref{fig:inf_vs_params} we show that repetition rates as low as $100-200$~MHz are compatible with high-fidelity pulse sequences in a long ion
chain ($N=60$). This significantly improves on the results of Ref.~\cite{Bentley2015}, where the authors reported the requirement of repetition rates ${\sim}5$ GHz to achieve high-fidelities in chains of $N>5$ ions. This is in part due to the differences in their model from ours. Their model took the axial trap frequency $\omega_t$ to be fixed as the ion chain increases in length, as opposed to our more experimentally realistic model where we reduce $\omega_t$ and instead fix $\omega_r$. Moreover, this improvement can largely be attributed to the superiority of our optimisation protocol which has more degrees of freedom and fewer constraints than previous schemes \cite{Gale2020}.

\section{\label{sec:exp_considerations}Experimental robustness}
\subsection{Pulse errors}
We have shown in this manuscript that errors of at least $\epsilon=10^{-3}$($10^{-4}$) are needed to achieve fidelities above $90\%$($99\%$). The requirement of high-accuracy $\pi$-pulses is an unfortunate consequence of the pulsed fast gate mechanism, and not specific to our model. To date, there have been only few experiments demonstrating single qubit control with ultra-fast pulses, with the state-of-the-art in pulse errors around $\epsilon=5\times10^{-3}$ \cite{Campbell2010,Wong-Campos2017,Heinrich2019}. These pulse errors are typically due to intensity fluctuations of the laser, which in the worst case (square pulses) give $\epsilon\approx \Delta I/I$. We note that our inclusion of pulse imperfections in the infidelity calculation is based on assumption that there are errors in each pulse that compound maximally. Thus the pulse errors for high-fidelity gates quoted above should be understood as worst-case requirements.  \par
We have previously commented \cite{Gale2020} that replacing each $\pi$ pulse with a robust composite-pulse sequence \cite{CompositePulses_Cummins} is a promising pathway to improving these errors. In particular, replacing each pulse with a four-pulse BB1 sequence that is insensitive to first and second order intensity fluctuations \cite{BB1_Wimperis} can dramatically improve the precision of the SDKs in the gate. However, this approach is only
robust to phase-fluctuations if the constituent pulses are split from a larger parent pulse. Typically ultra-fast pulses have durations on the order of tens of picoseconds, and require Rabi frequencies on the order of $100$~GHz \cite{Campbell2010}.  Generating multiple copies of this size from larger parent pulses will require approximately five times the power per pulse. This will prove experimentally challenging
, and it may be necessary to reduce laser repetition rate to dedicate increased laser power to creating larger pulses given the significant improvements the BB1 sequence promises.  In recent years, there has been a focus on improving laser repetition rates to enable pulsed fast gate schemes, however given that the gate solutions we present in this manuscript require low numbers of pulses and repetition rates ${\sim}100{-}300$MHz, it is appropriate the focus is shifted to improving the transfer probability of the ultra-fast $\pi$ pulses. Other prospects for improving these errors are the use of rapid-adiabatic passage \cite{Malinovsky2001}, and pulse-shaping methods \cite{Oksenhendler2010}. \par
We assume the counter-propagating $\pi$ pulses composing each SDK are split from the same parent pulse, in which case laser phase cancels exactly. Therefore, this fast gate model is completely insensitive to phase fluctuations
of the laser.

\subsection{Coulomb non-linearity} 
We have previously reported that the effect of Coulomb non-linearity for fast gates with $N_p\sim 1000$ SDKs is to introduce errors on the order of $10^{-6}$ in the fidelity \cite{Gale2020}. In this manuscript, we have considered a much lower momentum range, with no more than $N_p=100$ SDKs in any given gate (on average, $N_p{\sim}20$), and thus we expect Coulomb non-linearity to cause much smaller errors. Further, as realistic gate infidelity is dominated by pulse imperfections, errors due to Coulomb non-linearity can be neglected.  \par
\subsection{Ion shelving}
\begin{figure}[t!]
    \centering
    \includegraphics[width=\columnwidth]{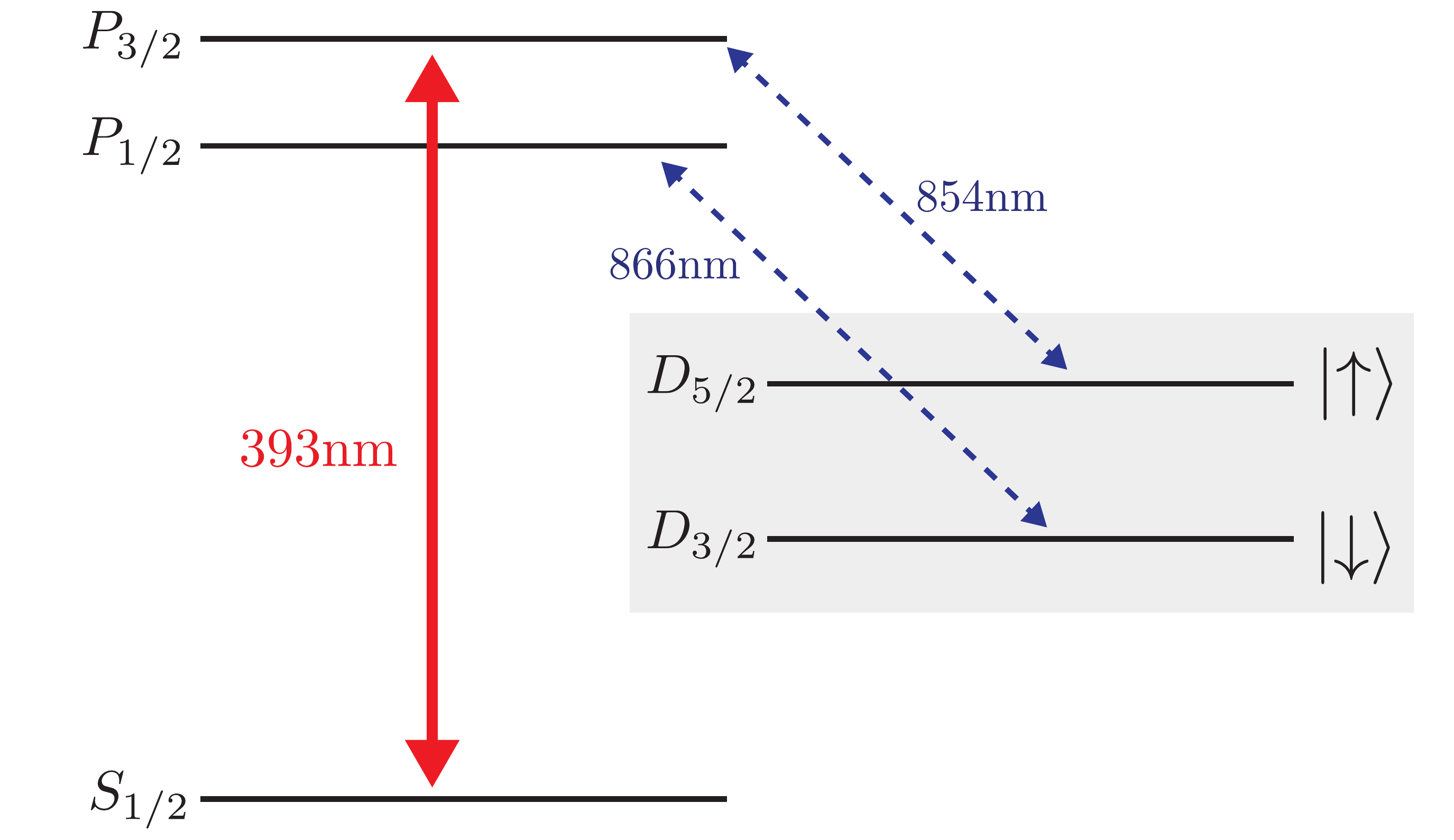}
    \caption{Simplified diagram of the electronic structure in $^{40}\text{Ca}^{+}$. The ultra-fast pulses are resonant with the $S_{1/2}\rightarrow P_{3/2}$ (red, solid), and the shelved qubits are encoded in the metastable $D_{3/2}$ and $D_{5/2}$ levels (shaded). Dipole transitions from the $D$ to $P$ levels are indicated (blue, dashed). Transition wavelengths are indicated, taken from \cite{NIST:ASD}. We note $D\leftrightarrow S$ dipole transitions are forbidden, and hence not indicated on this diagram.  }
    \label{fig:ShelvingDiagramCa40}
\end{figure}
Unwanted excitations can occur as a result of difficulties in addressing individual ions with the ultra-fast pulses.
This may be particularly challenging for ions in the center of the chain that are closer together than ions at the edge of the chain. We propose that these issues may be avoided by using long-wavelength (i.e. microwave) radiation with a global envelope to `shelve' all ions in the chain into internal states off-resonant with the ultra-fast laser pulses. The target ions for the gate can then be selectively `unshelved' with a pair of co-propagating radial Raman beams, where one beam covers all the ions with a global envelope and the other focuses on single ions, as in \cite{Debnath2016,Landsman2019}. The counter-propagating ultra-fast pulses can then be sent down the length of the ion chain to selectively interact with the targeted ions. The difficulty of focusing on single ions with the transverse Raman beam may be reduced by adding a quartic term to the potential to spread ions evenly throughout the chain, as in \cite{Lin2009,Pagano2018}. This is compatible with our optimisation protocol, which can design gates for arbitrary potentials so long as they are approximately harmonic around the equilibrium positions of the ions. 
\par
While this technique will protect the qubit state from being directly driven, the ultra-fast laser pulses may shift the energies of the shelved levels by the AC Stark effect. The main consequence of this is that the the two shelved levels may experience different shifts, which will result in a phase difference in the qubit. As a concrete example, we consider shelving of qubits into the $D_{3/2}$ and $D_{5/2}$ metastable levels in $^{40}\text{Ca}^{+}$ ions, as shown in Fig~\ref{fig:ShelvingDiagramCa40}. Assuming ultra-fast pulses 
resonant on the $S_{1/2}\rightarrow P_{3/2}$ transition with a Rabi frequency of $300$~GHz, we estimate the phase shift induced on the qubit from the AC Stark shift due to a single pulse to be approximately $6\times10^{-6}$ radians (see Appendix \ref{append:StarkShift}). The optimal gate sequences considered in this manuscript consist of roughly $10{-}30$ counter-propagating pulse pairs, the total phase shift will be no larger than about $10^{-4}$ radians. Therefore in this case, the phase shift is sufficiently small that it may be neglected, however in principle it may still be compensated for with precise single-qubit operations on each of the shelved ions. 

\subsection{Hot motional states} 
\begin{figure}
    \centering
    \includegraphics[width=\columnwidth]{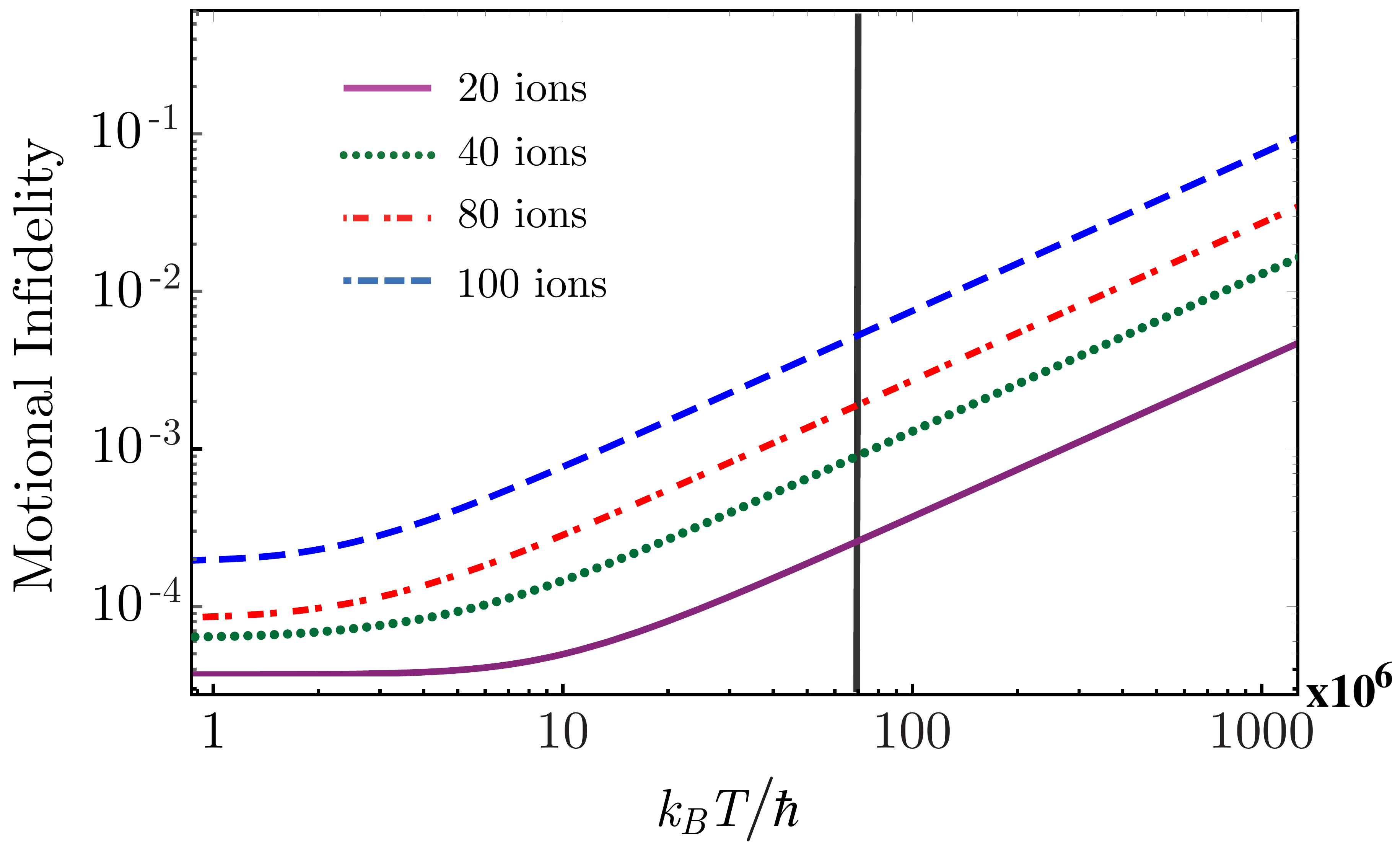}
    \caption{Motional infidelity (contribution of motional restoration errors to gate infidelity) as a function of the ions' temperature, for exemplary gates in ion chains of different length. The vertical line indicates the Doppler temperature $ k_B T_D/\hbar\approx 7\times10^7$~Hz.}
    \label{fig:MotInf_Temperature}
\end{figure}

This fast gate mechanism is insensitive to the initial motional state \cite{Garcia-Ripoll2003}. For the calculations in the manuscript we assume a thermal product state for each motional mode with an average phonon occupation of $\bar{n}=0.1$. For higher occupations, the contribution of the motional restoration terms to the infidelity will increase in magnitude (see the infidelity expressions given in Appendix \ref{Append:OptDetails}). However, we find that our optimised gates, even in long ion chains with weak axial trapping, are robust to higher temperature motional states, as shown in Fig~\ref{fig:MotInf_Temperature}. This can be explained by the coupling of gate dynamics to individual motional modes to the gate dynamics being weaker for gates designed for long ion chains, thus reducing the overall effect of larger mode occupations on the gate fidelity. \par
We consider cooling to the Doppler temperature utilising the $397\text{~nm }P_{1/2}\rightarrow S_{1/2}$ transition in $^{40}\text{Ca}^+$ ions, which gives $k_B T_D/\hbar=\Gamma_e/2\approx 7\times10^7$~Hz. As shown in Fig~\ref{fig:MotInf_Temperature} the gate error associated with this much higher temperature is less than $1\%$ even in $100$ ion chains with weak axial trapping. For smaller ion chains with $20$ ions, we find the error is on the order of $10^{-4}$. This enables the use of the axial modes which are typically more challenging to cool near ground state. Notably, pulsed fast gate schemes are valid outside of the Lamb-Dicke regime $\eta^2(2\Bar{n} +1)\ll 1$, and thus errors are not introduced by out-of-Lamb-Dicke effects which have dominated errors in continuous-pulse fast gates \cite{Schafer2018}.   \par 
As fast gates are robust to the ions' temperature, heating of the motional modes between gates will not degrade the fidelity of computation. However, a previous study by Taylor \textit{et al.} found that if a heating event occured during the gate operation, the gate fidelity is significantly damaged \cite{Taylor2017}. To minimise the probability of a heating event occuring during a gate operation, computation must be performed sufficiently fast with respect to the heating timescale. This effectively places a limit on the number of fast gates that can be performed in succession. However, as the gate speeds we have considered in this manuscript (${\sim}1~\mu$s) are orders of magnitude faster than typical trap heating rates, there is a high ceiling for the total number of entangling gates that can be performed without damage from trap heating. 
\par
\subsection{Misaligned beams} 
If the counter-propagating beams are not aligned correctly with the longitudinal axis of the trap, they may still be aligned with respect to each other such that their respective effect on the transverse ion motion will cancel. In this case, misalignment 
will result in an effective change of the magnitude of the momentum kicks, $\hbar\Delta k_\epsilon$. For systematic misalignment the motional trajectories will be restored by the end of the gate, leaving only an error in the enclosed phase-space area, resulting in a phase error that scales with ${\sim}\Delta k_\epsilon^2$ at worst. In the case where the counter-propagating beams are misaligned with respect to each other, there will be transverse modes coupled into the ion motion that will cause errors in both motional restoration and phase accumulation. Based on the expression for the infidelity given in Appendix \ref{Append:OptDetails}, we expect the infidelity to scale with the square of the magnitude of the transverse couplings. If this misalignment is systematic, it may be corrected for with an online optimisation on the experimental apparatus. 
\par
\subsection{Timing errors} 
The impact on gate fidelity due to Gaussian noise on SDK timings has previously been studied \cite{Ratcliffe2018}.  Here we discuss the effect of shot-to-shot fluctuations in the laser repetition rate and the trapping frequency, both of which change the SDK timings with respect to the motional timescale. We find that instabilities in both the trapping frequency or laser repetition rate of about $1\%$($0.1\%$) introduces errors of order $10^{-3}$($10^{-5}$) in the total gate fidelity. Given we have demonstrated pulse errors are a limiting factor, infidelity errors below ${\sim}10^{-3}$ can be neglected. This leaves requirements on trap frequency stability on the $0.1-1$~kHz level, and repetition rate stability on the MHz level. Both are well within the capability of current experiments; compatible high-repetition rate lasers can currently demonstrate fractional instabilities below roughly $10^{-7}$ \cite{Hussain2016,Heinrich2019}.

\section{Conclusions}
We have investigated a model for implementing fast two-qubit gates in long ion chains with ultra-fast pulses. By applying a two-stage protocol to find optimal pulse sequences demonstrated that high-fidelity ${\sim}1~\mu$s gates are achievable with low numbers of pulses repetition rate lasers within experimental capability. We have demonstrated the optimal gates are robust to a variety of experimental imperfections, with the exception of pulse area errors which have previously been identified as the dominant error source. We have found the experimental requirements for implementing high-fidelity gates are independent of the length of the ion chain. Provided the required pulse control can be achieved, this model may be used to realise large-scale quantum computation in long ion chains in near future experiments.

\begin{acknowledgments}
The authors thank Cornelius Hempel and Simon Haine for useful discussions. This research was undertaken with the assistance of resources and services from the National Computational Infrastructure (NCI), which is supported by the Australian Government. 
\end{acknowledgments}

\appendix
\pagebreak
\widetext
\section{\label{Append:OptDetails} Optimisation details}
For optimisation of the timings and numbers of the SDK sequence, we use the Anti-symmetric Pulse Group scheme \cite{Gale2020} with $N_k$ groups of SDKs,
\begin{align}
\begin{aligned}
    \mathbf{z} &= \left\{ -z_{N_k/2}, \, \dots, \, -z_2, \, -z_1, \, z_1, \, z_2, \, \dots, \, z_{N_k/2} \right\} \, , \\
    \mathbf{t} &= \left\{ -t_{N_k/2}, \, \dots, \, -t_2, \, -t_1, \, t_1, \, t_2, \, \dots, \, t_{N_k/2} \right\} \, .
\end{aligned}
\end{align}
where $z_j$ and $t_j$ are the number of SDKs in the $j^{th}$ SDK group and the time that group arrives at the ions. As described in \cite{Gale2020}, the anti-symmetry of $\mathbf{z}$ and $\mathbf{t}$ guarantees that the momentum of each motional mode is restored by the end of the SDK sequence. \par

We use the state-averaged infidelity of the unitary $U_\text{gate}$ defined in Equation (4) of the main text, with respect to the ideal unitary $U_\text{id}=\exp(i\frac{\pi}{4}\sigma_z^\mu\sigma_z^\nu)$, as a cost-function for optimisation of the pulse sequence. For these unitaries state-averaged infidelity is given by:
\begin{equation}
    1-F=1-\frac{1}{\int_{\psi_0}d\ket{\psi_0}}\int_{\psi_0}\textnormal{Tr}_m \left[ \bra{\psi_0}U^\dag_{id}U_{gate}(\ket{\psi_0}\bra{\psi_0}\otimes \rho_m )U^\dag_{gate} U_{id} \ket{\psi_0}\right] d\ket{\psi_0} \,,
\end{equation}
where we have taken the partial trace over the motional degrees of freedom, and the integrals are over the unit hypersphere of all two-qubit states. Assuming a thermal product state for $\rho_m$, and Taylor expanding for small errors in the restoration of each motional mode $\Delta \alpha_m$ and mismatch of the entangling phase $\Delta\phi$ the infidelity takes the approximate form:
\begin{equation}
    1-F \approx \frac{2}{3}|\Delta \phi| ^2 + \frac{4}{3}\sum_m (\frac{1}{2}+\Bar{n}_m) \big((b_m^\mu)^2+(b_m^\nu)^2 \big)|\Delta \alpha_m|^2
\end{equation}
where $\bar{n}_m$ is the mean phonon occupation of the $m$-th motional mode. This can be equivalently written in terms of ion temperature $T$ and mode frequencies $\omega_m$,
\begin{equation}
    1-F \approx \frac{2}{3}|\Delta \phi| ^2 + \frac{4}{3}\sum_m \text{coth}\left(\frac{\hbar \omega_m}{k_B T}\right) \big((b_m^\mu)^2+(b_m^\nu)^2 \big)|\Delta \alpha_m|^2 \,.
\end{equation}
In the main text we use the term `motional infidelity' to refer to the exclusion of the first term in the above expression. The expressions for the phase mismatch and motional restoration terms are given as:
\begin{align}
    \Delta \phi &= \bigg{|}8\eta_m^2 b_m^\mu b_m^\nu \sum_{i\neq j} z_i z_j \sin{(\omega_m |t_i-t_j|)}\bigg{|} -\frac{\pi}{4} \,, \\
    \Delta \alpha_m &= 2\eta_m \sum_{k = 1} z_k z_k \sin(\omega_m t_k) \,.    
\end{align}
Calculation of the mode coupling vectors $\mathbf{b_m}$ and frequencies $\omega_m$ can be calculated from the Hessian matrix, as described in the following section. \par
The first stage of the optimisation is over the elements $\{ z_1, \, z_2, \, \dots, \,z_{N_k/2}\}$, with the timings set at constant intervals $t_j = j\;T_G/N_k$. This optimisation is first performed with tight bounds on the number of SDKs in each group $|z_j|\leq 1$, and then for gradually loosened bounds up to $|z_j|\leq 10$. We find that the minimal number of pulse groups required for our optimisation to converge to high-fidelity solutions is $N_k=16$ for ions toward the edges of the chain, and $N_k=18$ for ions closer to the middle. Somewhat surprisingly, we find this to be independent of the length of the chain for $N\gtrsim 6$. For large ion chains $N\gtrsim 60$, we find optimal sequences often have $z_{N_k/2}=0$, reducing the total number of pulse groups in the sequence. This effectively shortens the gate time by roughly $10\%$. \par

The second stage takes optimal solutions for $\{z_j\}$ identified in the first stage, and performs sets of local optimisations on $\{ t_2-t_1, \, t_3-t_2, \, \dots, \,t_{N_k/2}-t_{N_k/2-1}\}$ on a grid of timings defined by a specified repetition rate. In these local optimisations, we allow the values of each $\{t_j-t_{j-1}\}$ to vary by up to $25\%$. This second stage utilises an ODE description of the gate dynamics (described in Appendix \ref{Append:ODES}), as opposed to using the expressions above for $\Delta\phi$ and $\alpha_m$, to include the effect of finite repetition rate separating SDKs in the same group. \par

\section{\label{Append:ODES}ODE description of gate dynamics in normal-mode basis}
The use of an ODE description of the ions motion to describe fast gate dynamics is described in depth in Ref.~\cite{Gale2020}. The key idea is that the motional state trajectory associated with each two-qubit basis state is well described by the classical motion of the trapped ions, and thus the gate dynamics can be modelled by, for each two-qubit state, evolving a set of ODEs with SDKs can be modelled as discontinuous transformations on the ions' velocities. In general these ODEs are non-linear due to the Coulomb interaction, which become computationally expensive to solve as the number of ions in the trap are increased. For the purposes of optimisation, we use linearisation to derive an ODE description that is numerically inexpensive to compute. \par
Here we derive the set of ODEs for ion motion in one-dimension, as in the main text we restrict our consideration to motion induced along the longitudinal axis of the trap.  We begin with a typical Lagrangian description of $N$ coordinates $\mathbf{x}=\{x_1(t),x_2(t),...,x_N(t)\}$:
\begin{equation}
    \mathcal{L}=\frac{1}{2}M\Dot{\mathbf{x}}\cdot\Dot{\mathbf{x}}-V(\mathbf{x},t) \,,
\end{equation}
where $V(\mathbf{x})$ describes both the (typically harmonic) trapping potential $V_\text{trap}(\mathbf{x})$ and the Coulomb interaction $V_\text{Coul}(\mathbf{x})=\sum_{j\neq k}\frac{q^2}{4\pi\epsilon_0|x_j-x_k|}$. Note that $V_\text{trap}(\mathbf{x})$ need not be harmonic, and can include higher order corrections, such as the stabilising quartic term mentioned in the main text. Linearisation is achieved by Taylor expanding the potential to second order about a set of equilibrium positions $\mathbf{x_0}$ (that satisfy $\frac{\partial V(\mathbf{x})}{\partial x_i}\big{|}_\mathbf{x_0} = 0$). This gives, up to a constant offset,
\begin{align}
    \mathcal{L} &= \frac{1}{2}M\Dot{\mathbf{x}}\cdot\Dot{\mathbf{x}} - \frac{1}{2}\sum_{i,j=1}^n \frac{\partial^2 V}{\partial x_i \partial x_j}\bigg{|}_\mathbf{x_0}(x_i(t)-x_0)(x_j(t)-x_0) +\dots \\
    &\approx \frac{1}{2}M\Dot{\mathbf{q}}\cdot\Dot{\mathbf{q}} -\frac{1}{2}M \mathbf{q}^T \cdot \underline{\mathbf{H}} \cdot\mathbf{q} \,,
\end{align}
where in the second line we have made a change of variables $\mathbf{q}(t)\equiv\mathbf{x}(t)-\mathbf{x_0}$ and defined the Hessian matrix $\underline{\mathbf{H}}$ which has components $H_{ij}=\frac{1}{M}\frac{\partial^2 V}{\partial x_i \partial x_j}\big{|}_\mathbf{x_0}$. The coupling vector $\mathbf{b_p}$ and the frequency $\omega_p$ of the $p$-th normal mode can be extracted from the Hessian by solving the eigenvalue equation
\begin{equation}
    \underline{\mathbf{H}}\cdot\mathbf{b_p} = \omega_p^2 \; \mathbf{b_p} \,.
\end{equation}
This linearised Lagrangian is sufficient to describe excursions of the ions positions about equilibrium. We now convert to the normal mode basis, where each mode is described by a coordinate $Q_p$, using the relation $\mathbf{q}=\sum_p \mathbf{b_p}Q_p$:
\begin{equation}
     \mathcal{L} = \frac{1}{2}M\sum_{p=1}^N \left( \Dot{Q}^2_p - \omega_p^2 Q_p^2 \right) \,.
\end{equation}
The corresponding Euler-Lagrange equations for this linearised Lagrangian are then given by a set of $N$ decoupled equations resembling those of simple harmonic oscillators:
\begin{equation}
    \Ddot{Q}_i(t) = - \omega_i^2 Q_i(t) \,,
\end{equation}
which can be easily solved numerically. The accumulated phase of the two-qubit state along the classical trajectory can be computed by taking the integral over the Lagrangian $\hbar\Phi = \int dt \mathcal{L}$, which we formulate as an ODE that can be evolved alongside the above equations:
\begin{equation}
    \hbar \Dot{\Phi} = \frac{1}{2}M\sum_{p=1}^N \left( \Dot{Q}^2_p - \omega_p^2 Q_p^2 \right) \,.
\end{equation}

\section{\label{append:StarkShift} Stark shifts on a shelved qubit in $^{40}\text{Ca}^{+}$.}

As an illustrative example of the ion shelving discussed in text, we consider a qubit `shelved' in the metastable $D_{3/2}$ and $D_{5/2}$ electronic states of $^{40}\text{Ca}^{+}$. We assume ultra-fast $\pi$ laser pulses resonant on the $\SI{393}{\nano\meter}$ $S_{1/2}\rightarrow P_{3/2}$ transition, as in \cite{Heinrich2019}, with constant Rabi frequency $\Omega_{S_{1/2}\rightarrow P_{3/2}}=300$~GHz for simplicity. A simplified diagram of the electronic scheme and relevant transitions is shown in Fig~\ref{fig:ShelvingDiagramCa40}. As the $S\rightarrow D$ transitions in $^{40}\text{Ca}^{+}$ are dipole-forbidden, the leading order Stark shift on the $D_{3/2}$ state is given by:
\begin{equation}
    \Delta E_\text{Stark}^{D_{3/2}} \approx \frac{|\Omega_{D_{3/2}\rightarrow P_{1/2}}|^2 }{4|\omega_L - \omega_{D_{3/2}\rightarrow P_{1/2}}|}.
\end{equation}
Using data given in \cite{NIST:ASD} to calculate the transition dipole moments $\mu_{ij}$, we find 
\begin{equation}
    \Omega_{D_{3/2}\rightarrow P_{1/2}}=\frac{\mu_{D_{3/2},P_{1/2}} }{\mu_{S_{1/2},P_{3/2}}}\Omega_{S_{1/2}\rightarrow P_{3/2}}\approx 0.88\Omega_{S_{1/2}\rightarrow P_{3/2}} \,,
\end{equation} 
and thus calculate $ \Delta E_\text{Stark}^{D_{3/2}}/\hbar\approx 6.62\times10^6\text{~s}^{-1}$. Similarly, we find $\Delta E_\text{Stark}^{D_{5/2}}/\hbar\approx 6.01\times10^6\text{~s}^{-1}$. We can calculate the duration of each pulse from the requirement that $\tau_\pi\times\Omega_{S_{1/2}\rightarrow P_{3/2}}=\pi$, giving $\tau_\pi\approx\SI{10.5}{picoseconds}$. The phase shift between the two metastable levels from the Stark shift due to a single ultra-fast pulse can then be calculated as:
\begin{equation}
    \Delta\phi_\text{Stark} = \frac{\tau_\pi}{\hbar}\left(\Delta E_\text{Stark}^{D_{5/2}}-\Delta E_\text{Stark}^{D_{3/2}}\right)\approx 6.51\times10^{-6} \text{~radians} \,.
\end{equation}

\bibliography{bibliography}

\end{document}